\documentclass[lettersize,journal]{IEEEtran}
\usepackage{amsmath,amsfonts}
\usepackage{algorithmic}
\usepackage{algorithm}
\usepackage{array}

\usepackage[caption=false,font=normalsize,labelfont=sf,textfont=sf]{subfig}
\usepackage{textcomp}
\usepackage{stfloats}
\usepackage{url}
\usepackage{verbatim}
\usepackage{graphicx}
\usepackage{amssymb}
\usepackage{cite}
\usepackage{booktabs} %

\usepackage[colorlinks,linkcolor=blue,citecolor=blue]{hyperref}
\IEEEoverridecommandlockouts
\renewcommand{\IEEEpubid}{}

\usepackage{tikz,xcolor}

\newcommand{\orcidicon}{%
	\begin{tikzpicture}
		\draw[lime, fill=lime] (0,0)
		circle[radius=0.16]
		node[white]{{\fontfamily{qag}\selectfont \tiny \.{I}D}};
	\end{tikzpicture}%
}
\newcommand{\orcidauthorA}{\href{https://orcid.org/0000-0002-1182-4035}{\orcidicon}}
\newcommand{\orcidauthorB}{\href{https://orcid.org/0009-0001-4165-1630}{\orcidicon}}

\begin{document}

\title{Convolution-Augmented Transformer for Robust Blind CFO Estimation in LEO Satellite Communication Systems}

\author{Miaomiao Wei\orcidauthorA, Huaiyuan Chen\orcidauthorB, Xinnan Shao, Lulu Gao, Hang Liu%
	\thanks{This work was supported in part by the National Natural Science Foundation of China under Grant 62301624, in part by the Scientific Research Projects of Higher Education Institutions of Henan Province under Grant 23A510012, in part by the Key Laboratory of Water Management and Water Security for Yellow River Basin, Ministry of Water Resources, under Grant 2026-SYSJJ-07, and in part by the Science and Technology Guiding Plan Project of China National Textile and Apparel Council under Grant 2026005. \textit{(Corresponding author: Miaomiao Wei.)}}%
	\thanks{Miaomiao Wei, Huaiyuan Chen, Xinnan Shao, Lulu Gao, and Hang Liu are with the School of Information and Communication Engineering, Zhongyuan University of Technology, Zhengzhou 450001, China (e-mail: \mbox{mmwei@zut.edu.cn}; \mbox{hychy031@163.com}).}%
}

\markboth{}{}

\maketitle

\begin{abstract}
Low Earth Orbit (LEO) satellite communications are essential for achieving ubiquitous global connectivity. However, the high mobility of LEO satellites induces massive Doppler shifts and fast time-varying fading, severely impairing signal synchronization. Accurate Carrier Frequency Offset (CFO) estimation is critical for receiver synchronization. Nevertheless, conventional estimators and deep learning approaches often degrade significantly under highly dynamic, low-SNR conditions. To address these challenges, this paper proposes a novel Convolution-Augmented Transformer architecture for robust, blind CFO estimation in LEO communication systems. Initially, to overcome the inherent limitations of standard linear embedding layers in capturing nonlinear phase rotations, we construct a multi-domain input feature that encompasses in-phase/quadrature (I/Q), polar, and statistical autocorrelation properties. Subsequently, a Convolutional Pre-Denoising (CPD) module, integrating multi-scale dilated convolutions and an attention mechanism, is deployed at the front end to adaptively suppress localized noise and extract robust spatiotemporal representations. These refined latent features are then processed by a Transformer encoder to explicitly track global, long-range phase dependencies. Extensive simulation results demonstrate that the proposed data-driven approach significantly outperforms conventional estimation algorithms, achieving enhanced estimation accuracy and superior resilience in highly dynamic LEO environments.

\end{abstract}


\section{INTRODUCTION}
\IEEEPARstart{D}RIVEN by the escalating demand for higher transmission speeds, lower latency, and enhanced reliability, communication system architectures have continuously evolved, leading to significant advancements in wireless communication technology over the past few decades. Nevertheless, the expansion of terrestrial networks remains constrained by fundamental limitations, particularly in remote and underserved regions. In such areas, terrestrial infrastructure fails to provide seamless coverage, as deployment challenges and prohibitive costs are most acute \cite{zhao2024qos,chen2024spaceedge}. Non-terrestrial networks (NTNs) serve as an essential solution for realizing the ubiquitous connectivity envisioned for sixth-generation (6G) systems, particularly in regions devoid of terrestrial infrastructure\cite{ntnadd}. Within this paradigm, satellite communication systems, specifically those employing low Earth orbit (LEO) satellites, have emerged as a highly promising solution. By offering low latency, high throughput, and global coverage, LEO satellites possess the potential to transform existing communication paradigms and deliver reliable connectivity to regions inaccessible to traditional infrastructure \cite{kodheli2020satellite}.

To meet the high-data-rate requirements of future satellite services, Orthogonal Frequency Division Multiplexing (OFDM) is considered a core modulation scheme for LEO systems owing to its high spectral efficiency and robustness against multipath fading. However, integrating OFDM into LEO satellite links introduces significant technical hurdles, primarily stemming from the high mobility of the platform \cite{Tomeet}. LEO satellites operate in orbits relatively close to the surface of the Earth (approximately 500 to 2000 km), traveling at orbital velocities exceeding 7.5 km/s. This rapid relative motion between the satellite and the ground terminal induces massive Doppler shifts and time-varying frequency offsets\cite{ntnadd3}.

Although OFDM fundamentally relies on maintaining strict orthogonality among subcarriers, the severe Doppler effect in LEO scenarios induces a frequency-domain shift within the signal spectrum. This spectral misalignment triggers Inter-Carrier Interference (ICI) and profound synchronization errors \cite{ICI},\cite{iciadd}. Unlike traditional terrestrial networks, LEO systems are characterized by pronounced Doppler dynamics, featuring both large frequency offsets and rapid temporal variations\cite{ntnadd2}. Together, these factors create a fast time-varying channel that rapidly degrades the signal-to-interference-plus-noise ratio (SINR). The throughput gains offered by OFDM can be realized only when the dynamic carrier frequency offset (CFO) is accurately estimated and compensated. Robust frequency offset synchronization is therefore essential to reliable transmission over LEO satellite links.

Receiver synchronization requires precise estimation and compensation of the carrier frequency offset. Conventional estimation techniques typically rely on accurate signal models and rigorous mathematical derivations. They estimate the frequency offset using analytical estimators derived from the signal structure. In signal processing, these approaches are commonly grouped into time-domain and frequency-domain schemes. Time-domain schemes make use of the phase relationships and autocorrelation characteristics of the received signal. Classical approaches, including the Fitz \cite{Fitz}, L\&R \cite{LR}, and M\&M \cite{MandM} estimators, utilize the phase difference between adjacent samples for frequency estimation. Recently, several improved time-domain methods have emerged. For instance, cyclic prefix-based estimation techniques \cite{cp-1,cp-2} exploit the correlation between the cyclic prefix and the tail of the OFDM symbol to estimate the carrier frequency offset without pilot overhead. Some studies design structured training sequences \cite{trainseq} or repetitive subsymbols \cite{rept-sym} to enhance estimation accuracy and extend the effective estimation range. However, numerous time-domain estimators depend on specially designed training sequences, which introduce additional pilot overhead and consequently degrade spectral efficiency. By contrast, frequency-domain approaches typically apply the discrete Fourier transform to form the periodogram of the received signal and then determine the frequency offset from the location of the spectral peak \cite{dft-1,dft-2,dft-3}. Nevertheless, approaches based on the discrete Fourier transform suffer from inherent limitations, such as the picket-fence effect and spectral leakage, which can introduce significant estimation errors. Furthermore, a limited observation length constrains the frequency resolution of the periodogram \cite{limit-dft}. To alleviate these issues, various interpolation-based techniques have been proposed to refine the coarse estimates obtained from the spectrum. For example, techniques utilizing parabolic or multi-point interpolation exploit the spectral peak and its adjacent spectral bins to derive a correction term, thereby refining the initial coarse frequency estimate \cite{interpolation-1,interpolation-2}. Nonetheless, these methods inherently depend on accurate coarse estimates, and their performance typically degrades under low SNR conditions or limited observation lengths.

Deep learning (DL) has recently attracted considerable attention in wireless communication research \cite{DLintro,DLintro2,DLintro3} owing to its strong capability in capturing nonlinear relationships within wireless signals. It has been successfully applied to various signal processing tasks, including signal source number estimation, channel estimation, signal identification, and blind signal separation \cite{DLeg1,DLeg2,DLeg3,DLeg4}. Building upon these successes, the exploration of DL for carrier frequency offset (CFO) estimation has garnered increasing interest. Early studies predominantly relied on convolutional neural networks (CNNs) \cite{cfoeg1,CNNeg-1,CNNeg-2} to extract representative features from received signals. However, approaches based on CNNs focus primarily on local feature extraction and exhibit limited capability in modeling the temporal dependencies of communication signals. To better capture the temporal characteristics of received signals, recurrent neural networks (RNNs) and associated variants, such as long short-term memory (LSTM) and gated recurrent unit (GRU) architectures, have been introduced for CFO estimation \cite{rnn-1,rnn-2}. For instance, Feng et al. \cite{rnn-2} propose an end-to-end framework that combines residual CNNs, LSTMs, and GRUs. This architecture facilitates a progressive workflow comprising spatial-frequency feature extraction, multi-user separation, and joint frequency offset estimation. Furthermore, the proposed model effectively mitigates the vanishing gradient problem while optimizing both parameter efficiency and computational complexity. More recently, architectures based on Transformers have attracted increasing attention in wireless signal processing \cite{transtormer} owing to the powerful self-attention mechanism, which enables the effective modeling of long-range dependencies within communication signals. For example, receivers utilizing Transformers have been proposed for OFDM systems to jointly learn channel characteristics \cite{ofdm-trans} and perform signal detection tasks \cite{sig-det}. These studies demonstrate the potential of Transformer networks in modeling complex structures of wireless signals and further inspire the application of these networks to synchronization problems, such as CFO estimation. Overall, existing literature indicates that deep learning provides a promising framework for CFO estimation by enabling data-driven feature extraction and the nonlinear modeling of communication signals.

This paper develops a robust blind CFO estimation framework for LEO communication systems under rapidly varying Doppler shifts, with multipath effects taken into account. The main contributions are summarized as follows:

\begin{itemize}
	\item \textbf{A Novel Convolution-Augmented Transformer Architecture:} We propose a deep learning-based method for blind CFO estimation that operates without requiring prior knowledge, such as training sequences or cyclic prefixes. Equipped with a Convolutional Pre-Denoising (CPD) module at the front end, the proposed architecture adaptively suppresses local high-frequency noise, demonstrating superior robustness even under low SNR conditions.
	
	\item \textbf{Multi-Domain Feature Construction:} To overcome the inherent limitations of standard linear embedding layers in directly capturing nonlinear phase rotations from raw in-phase and quadrature (I/Q) data, we design a multi-domain feature extraction scheme. By explicitly augmenting the input of the network with polar attributes (magnitude and phase) alongside statistical autocorrelation properties, we provide highly complementary information. This explicitly guided fusion of physical features significantly reduces the difficulty of learning dynamic Doppler shifts for the self-attention mechanism.
				
	\item \textbf{Comprehensive Performance Validation:} We conduct extensive ablation studies and evaluate the performance of the proposed method under conditions typical of LEO systems. Simulation results demonstrate that this data-driven approach significantly outperforms conventional algorithms by achieving a substantially lower mean squared error (MSE) in CFO estimation, thereby verifying the effectiveness and robustness of the proposed architecture.
			
\end{itemize}

The remainder of this paper is organized as follows. Section II introduces the system model. Section III presents the network architecture of the proposed estimation method and details the corresponding implementation. Section IV provides an analysis of the simulation results, and Section V concludes the paper.

\begin{figure}[!t]
	\centering
	\includegraphics[width=3.5in]{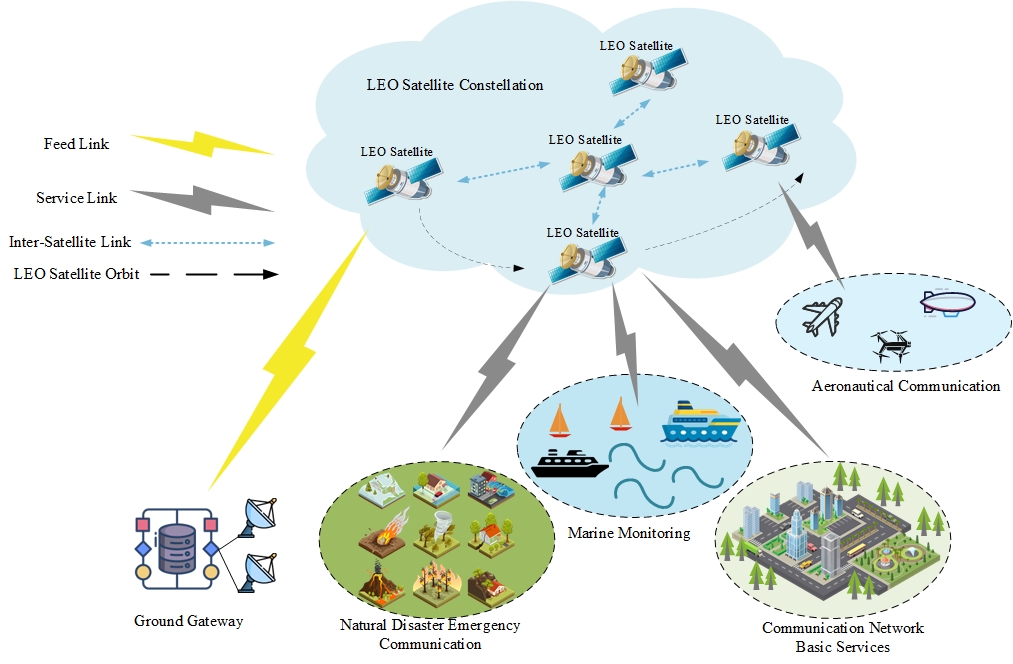}
	\caption{LEO satellite communication system.}
	\label{fig1}
\end{figure}

\section{SYSTEM MODEL AND PROBLEM FORMULATION}
As a critical interface of space-ground integrated networks, satellite-to-ground links enable wide-area coverage for terrestrial users. The considered communication architecture is illustrated in Fig. \ref{fig1}.  When the user equipment (UE) is located in areas with dense terrestrial infrastructure, such as urban regions with high-density deployments of small cells, communication is predominantly supported by terrestrial networks, whereas low Earth orbit (LEO) satellites provide supplementary capacity. In contrast, in rural or remote areas beyond the coverage of terrestrial cellular systems, LEO satellite links play a pivotal role in delivering reliable connectivity. For subsequent signal processing, we consider an OFDM-based satellite communication system that employs the transparent payload architecture defined by 3GPP \cite{3GPP_TS38211}. Within this architecture, the gateway station generates OFDM symbols and transmits them to the LEO satellite via the feeder link. The satellite operates in a transparent relay mode, where the received signal is amplified and forwarded to the user terminal through the service link.
\vspace {-\baselineskip}
\subsection{LEO Orbital Geometry Model}
Compared to terrestrial networks, satellite communication systems operate in more complex environments. Large-scale fading within satellite-to-ground links comprises three components: basic path loss, atmospheric loss, and scintillation loss. The total path loss is formulated as:
\begin{equation}
	PL = PL_b + PL_g + PL_s
	\label{eq:pl_calculation} 
\end{equation}

where $PL$ represents the total path loss, $PL_b$ denotes the basic path loss, $PL_g$ indicates the atmospheric loss, and $PL_s$ stands for the scintillation loss. Specifically, $PL_b$ is expressed as:
\begin{equation}
	PL_b = FSPL(d, f_c) + SF + CL
	\label{eq:pl_b}
\end{equation}

where $FSPL(d, f_c) = 32.45 + 20\log_{10}(f_c) + 20\log_{10}(d)$ represents the free space path loss, $d$ represents the distance from the satellite to the ground base station, $f_c$ denotes the carrier frequency, $SF$ indicates shadow fading, and $CL$ stands for the loss induced by ground objects. Based on the aforementioned fading characteristics, the overall channel is modeled as:
\begin{equation}
	h(t) = 10^{-\frac{PL}{20}} \bar{h}(t)
	\label{eq:channel}
\end{equation}

where $\bar{h}(t)$ denotes the small-scale fading component, and the factor $10^{-PL/20}$ converts the path loss in dB into a linear amplitude attenuation factor.

In practical LEO satellite communication scenarios, the high altitude of the satellite typically ensures the presence of a dominant direct propagation path, particularly at high elevation angles. Nevertheless, scattering and reflection caused by ground obstacles still contribute to multipath propagation. To accurately characterize this propagation environment, the small-scale fading component $\bar{h}(t)$ is decomposed into a deterministic line-of-sight (LoS) component and a stochastic non-line-of-sight (NLoS) component ($ \bar{h}(t) = h^{\text{LoS}}(t) + h^{\text{NLoS}}(t) $), where $ h^{\text{LoS}}(t) $ and $ h^{\text{NLoS}}(t) $ represent the LoS and NLoS channel components.

\vspace {-\baselineskip}

\subsection{Doppler Frequency Offset Model}
\begin{figure}[!t] %
	\centering
	\includegraphics[width=3.5in]{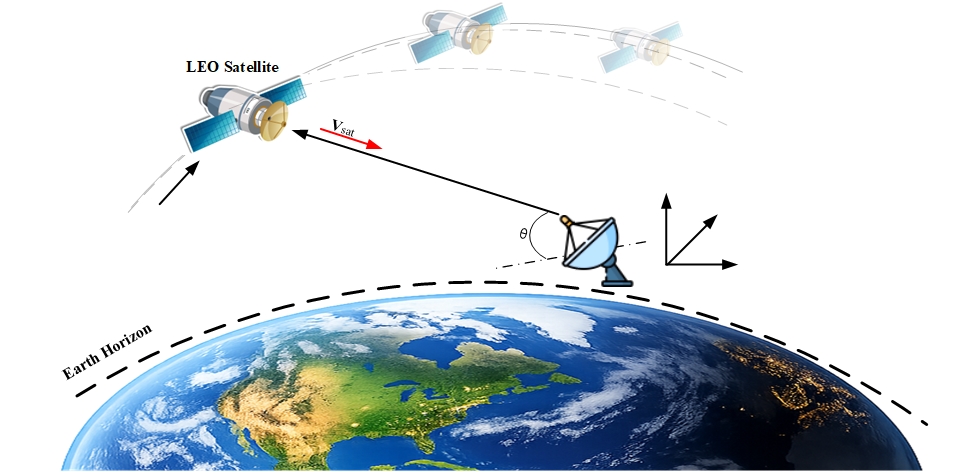}
	\caption{Doppler frequency shift.}
	\label{fig2}
\end{figure}

In LEO satellite links, the Doppler offset depends mainly on the carrier frequency and the relative velocity between the satellite and the ground terminal. Its time derivative, referred to as the Doppler rate, characterizes how rapidly the frequency shift changes and is therefore important for synchronization. As shown in Fig. \ref{fig2}, the high orbital speed of LEO satellites produces much larger Doppler offsets and Doppler rates than those in terrestrial scenarios. The Doppler offset can be expressed as \cite{doppler}:

\begin{equation}
	f_{d}(t) = \frac{v_{sat}}{c} f_c \cos \theta(t)
	\label{eq:doppler_max}
\end{equation}

where $f_c$ is the carrier frequency, $v_{sat}$ denotes the relative velocity between the satellite and the ground terminal, $\theta(t)$ denotes the angle between the satellite velocity vector and the satellite-to-ground LOS, and $c$ is the speed of light. Accordingly, the Doppler rate is defined as the time derivative of $f_d(t)$, i.e., $\dot{f}_d(t)=\frac{d f_d(t)}{dt}$. In high speed LEO scenarios, when the satellite is at a low elevation angle during an access period, the velocity projection onto the LOS becomes large, thereby resulting in a pronounced Doppler shift.

\vspace {-\baselineskip}
 
\subsection{OFDM Signal Model in LEO Systems}
Information is mapped onto Quadrature Amplitude Modulation (QAM) or Phase-Shift Keying (PSK) symbols over a set of orthogonal sinusoidal signals. We consider an OFDM system with $N$ subcarriers. Let $X_{l,k}$ denote the complex data symbol carried by the $k$-th subcarrier in the $l$-th OFDM symbol, where $k \in \left[0 , N-1\right]$ and $l \in \mathbb{Z}$. The corresponding time-domain OFDM symbol is obtained through the inverse Fourier transform as
\begin{equation}
	\label{deqn:signal_original}
	x_l(n) = \frac{1}{N} \sum_{k=0}^{N-1} X_{l,k} e^{j2\pi \frac{kn}{N}}, \quad 0 \leq n \leq N-1
\end{equation}

After the addition of a cyclic prefix (CP) of length $N_{\text{CP}}$, the extended $l$-th OFDM symbol is expressed as:
\begin{equation}
	\label{eq:signal_add_cp}
	x_l(n) =
	\begin{cases}
		x_l(n + N - N_{\text{CP}}), & 0 \leq n \leq N_{\text{CP}} - 1 \\
		x_l(n - N_{\text{CP}}), & N_{\text{CP}} \leq n \leq N_T - 1
	\end{cases}
\end{equation}

where $N_T$ denotes the total length of the discrete-time OFDM symbol after CP insertion, i.e., $N_T = N + N_{\text{CP}}$. To derive the frequency-domain input-output relationship, a symbol-wise quasi-static approximation is adopted, under which the channel is treated as constant over one OFDM symbol. In addition, the cyclic prefix is assumed to satisfy $N_{\text{CP}} \ge L - 1$, so that after CP removal the linear convolution with the channel is converted into a circular convolution over the useful symbol interval and inter-symbol interference is avoided. Ideal symbol timing synchronization is assumed, so that the receiver window is aligned with the useful symbol duration and timing offset is neglected. Using the discrete-time equivalent form of \eqref{eq:channel}, let $h_l[\tau]$, $0 \le \tau \le L-1$, denote the $L$-tap sampled channel impulse response (CIR) during the $l$-th OFDM symbol. The received signal can then be written as:
\begin{equation}
	\label{deqn:signal_receive_key}
	r_l[n] = \left( x_l[n] \ast h_l[n] \right) e^{j\frac{2\pi \varepsilon n}{N}} + z_l[n]
\end{equation}

where $\ast$ denotes linear convolution, $z_l[n]$ denotes additive white Gaussian noise (AWGN), and $\varepsilon$ represents the normalized carrier frequency offset. To facilitate the subsequent derivation, the frequency-domain equivalent form of  \eqref{deqn:signal_receive_key} is considered, in which the time-domain convolution is represented as multiplication in the frequency domain. Specifically, the transmitted frequency-domain signal $X_l[k]$ is multiplied by the channel frequency response $H_l[k]$, transformed back to the time domain, and then subjected to the phase rotation induced by the frequency offset. After removing the CP and retaining the $N$ useful samples, the equivalent time-domain signal can be expressed as
\begin{equation}
	\label{deqn:signal_receive_frequencydomain}
	y_l[n] = \left[ \frac{1}{N} \sum_{p=0}^{N-1} X_l[p] H_l[p] e^{j\frac{2\pi p n}{N}} \right] e^{j\frac{2\pi \varepsilon n}{N}} + z_l[n]
\end{equation}

Note that the summation index is denoted by $p$ to distinguish it from the index $k$ used in the Discrete Fourier Transform (DFT) at the receiver. Applying the DFT to the received time-domain signal $y_l[n]$ yields the following frequency-domain representation:
\begin{equation}
	\label{deqn:signal_receive_fft_process}
	Y_l[k] = \operatorname{DFT}\{y_l[n]\} = \sum_{n=0}^{N-1} y_l[n] e^{-j\frac{2\pi kn}{N}}
\end{equation}

From (\ref{deqn:signal_receive_frequencydomain}) and (\ref{deqn:signal_receive_fft_process}), the following formulation can be derived:

\begin{equation}
	\label{deqn:receive_signal_after_fft}
	\begin{split}
		Y_l[k]
		&= \sum_{n=0}^{N-1} \left(
		\frac{1}{N} \sum_{p=0}^{N-1} X_l[p] H_l[p]
		e^{j\frac{2\pi(p+\varepsilon)n}{N}}
		+ z_l[n]
		\right) \\
		&\quad \cdot e^{-j\frac{2\pi k n}{N}}
	\end{split}
\end{equation}

Simplifying (\ref{deqn:receive_signal_after_fft}) yields the closed-form expression:
\begin{equation}
	\label{deqn:receive_signal_after_fft_simple}
	\begin{split}
	Y_l[k]
	&= \frac{\sin(\pi\varepsilon)}{N\sin\left(\frac{\pi\varepsilon}{N}\right)} e^{j\pi\varepsilon \frac{N-1}{N}} X_l[k]H_l[k] \\
	&\quad + \sum_{\substack{p=0 \\ p\neq k}}^{N-1} X_l[p]H_l[p] \frac{\sin\left[\pi(p+\varepsilon-k)\right]}{N\sin\left[\frac{\pi(p+\varepsilon-k)}{N}\right]} \\
	&\quad \cdot e^{j\pi\frac{(p+\varepsilon-k)(N-1)}{N}} + Z_l[k]%
	\end{split}
\end{equation}

In~(\ref{deqn:receive_signal_after_fft_simple}), the first term represents the desired signal component affected by the residual CFO, where $\frac{\sin(\pi\varepsilon)}{N\sin\!\left(\frac{\pi\varepsilon}{N}\right)}$ denotes the amplitude attenuation and
$e^{j\pi\varepsilon\frac{N-1}{N}}$
denotes the phase rotation.
The second term represents the inter-carrier interference (ICI) caused by the residual frequency offset among subcarriers.
Moreover, $Z_l[k]$ denotes the corresponding frequency-domain noise term, which is assumed to follow an independent and identically distributed (i.i.d.) complex Gaussian distribution, i.e., $\mathcal{CN}(0,N_0)$.

\begin{figure*}[!t]  %
	\centering  %
	\includegraphics[width=7in]{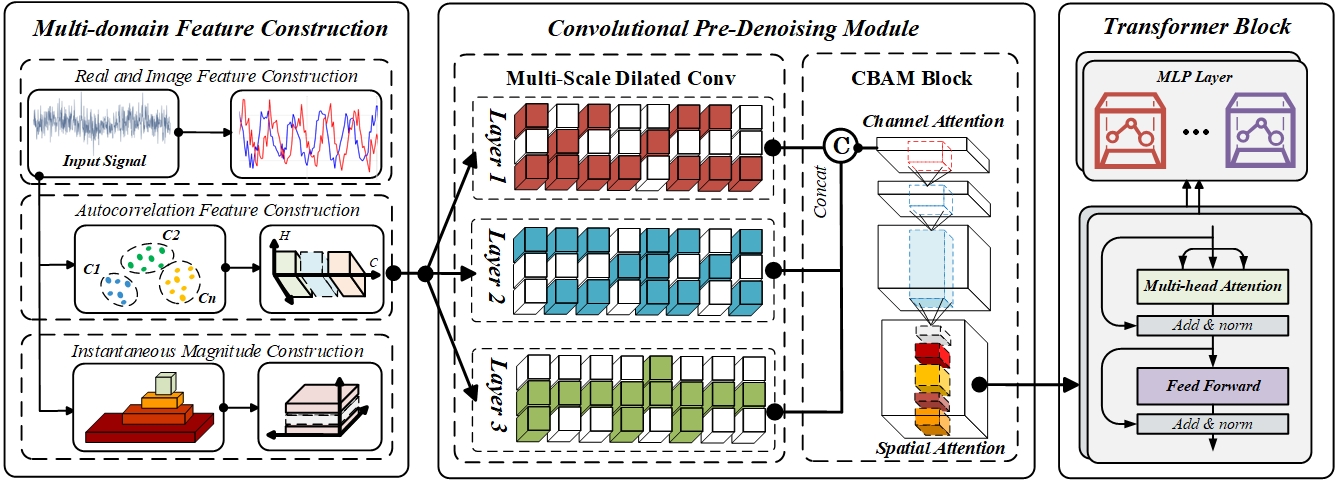}  
	\caption{Overall Architecture of the Proposed Blind CFO Estimation Network for LEO Communication Systems.}
	\label{fig:network_model}
\end{figure*}

\section{CARRIER FREQUENCY OFFSET ESTIMATION METHOD}

Due to the highly dynamic nature of LEO satellites, the resulting Doppler frequency shift is particularly severe. Fig. \ref{fig:network_model} illustrates the deep-learning framework developed for CFO estimation in LEO communication systems. The design is motivated by the large and rapidly time-varying Doppler shifts induced by the high mobility of LEO satellites. Under these conditions, conventional CFO estimators based on cyclic prefixes or training sequences often exhibit a limited estimation range and degraded accuracy. This section first presents the overall network pipeline, then details the implementation of each component, and finally analyzes the computational complexity of the proposed method.

\vspace {-\baselineskip}
\subsection{Overall pipeline}
The proposed framework follows a hierarchical feature learning paradigm. The Convolutional Pre-Denoising (CPD) module is first employed to capture local temporal patterns and suppress noise interference. Subsequently, the Transformer encoder, integrated with Rotary Positional Encoding (RoPE), explicitly models the long-range dependencies of phase evolution induced by the carrier frequency offset. Finally, an Attention Pooling mechanism is utilized to dynamically evaluate the reliability of each temporal feature, thereby generating a robust, noise-resistant representation for the ultimate CFO regression.

\vspace {-\baselineskip}
\subsection{Multi-domain Feature Construction}
To address the limitations of the methods for CFO estimation described in Section II (such as the reliance on prior knowledge, sensitivity to low SNR, and insufficient feature extraction), we initially perform preprocessing on the input signal. Because the received signal is complex in nature, Equation (\ref{deqn:signal_receive_key}) can be rewritten as:
\begin{equation}
	\label{sec3_IQ}
	y(n) = I(n) + Q(n)\mathrm{i}
\end{equation}

Although the $(I, Q)$ components theoretically contain all necessary information, the linear layers within the Transformer architecture encounter difficulties in directly capturing nonlinear phase rotations. The explicit provision of phase information can significantly facilitate the learning of Doppler shifts (which act as rotation operators) by the model. The specific calculation is formulated as:
\begin{equation}
	\label{sec3_phase}
	y_{\mathrm{phase}}(n) = \operatorname{atan2}\!\left(\Im\{y(n)\}, \Re\{y(n)\}\right)
\end{equation}
where $\Re\{y(n)\}$ and $\Im\{y(n)\}$ denote the in phase and quadrature components of the signal, respectively.

In traditional signal processing, the estimation of frequency deviation (such as the L\&R algorithm \cite{LR} and the M\&M algorithm \cite{MandM}) fundamentally utilizes the phase of the autocorrelation function. This function exhibits robust resistance against noise, and the position of the corresponding peak is directly correlated with the frequency deviation. The incorporation of autocorrelation features enables the attention mechanism within the Transformer to identify periodic deviations within the signal more readily. Additionally, considering the variation in the envelope of the complex signal $y[n]$ under a fading channel, the instantaneous magnitude feature $y_{\text{mag}}[n]$ is extracted. This feature is defined as the Euclidean distance from the origin to the signal within the complex plane:
\begin{equation}
	\label{sec3_mag}
	y_{\mathrm{mag}}(n) = |y(n)| = \sqrt{\Re\{y(n)\}^2 + \Im\{y(n)\}^2}
\end{equation}
\begin{equation}
	\label{sec3_auto}
	R_{\mathrm{raw}}[m] = \mathrm{IFFT}\{Y[k]Y^*[k]\}
	= \frac{1}{N_s}\sum_{n=0}^{N_s-1} y[n]\, y^*\!\left[(n-m)_{N_s}\right]
\end{equation}

where $Y^*[k]$ denotes the complex conjugate of $Y[k]$, and $(n-m)_{N_s}$ denotes the circular shift operation with modulus $N_s$, where $N_s$ is the length of the observation sequence used for autocorrelation computation. To eliminate the influence of signal power, the raw autocorrelation is normalized by its zero-lag value.
\begin{equation}
	\label{sec3_auto_1}
	R_{\mathrm{norm}}[m] = \frac{R_{\text{raw}}[m]}{R_{\text{raw}}[0]}
\end{equation}
where $R_{\mathrm{raw}}[0]$ corresponds to the total signal energy. For feature construction, the normalized autocorrelation sequence is aligned with the input index and decomposed into its real and imaginary parts, i.e.,$y_{\text{Corr\_Re}}[n] = \Re\{R_{\mathrm{norm}}[n]\}$, and $y_{\text{Corr\_Im}}[n] = \Im\{R_{\mathrm{norm}}[n]\}$. Finally, the input feature vector of the network at each time step $n$, denoted by $\boldsymbol{\xi}_n \in \mathbb{R}^6$, is constructed by concatenating the instantaneous physical attributes and the statistical correlation properties
\begin{equation}
	\label{input}
	\mathbf{\xi}_n = \left[y_{\text{Re}}, y_{\text{Im}}, y_{\text{mag}}, y_{\text{phase}}, y_{\text{Corr\_Re}}, y_{\text{Corr\_Im}}\right]^T
\end{equation}

where $y_{\mathrm{Re}}[n]=I(n)$ and $y_{\mathrm{Im}}[n]=Q(n)$. Specifically, the inclusion of normalized autocorrelation provides a noise-resilient representation of the temporal correlation properties inherent in the propagation environment.
\begin{figure}[!t] %
	\centering
	\includegraphics[width=3.5in]{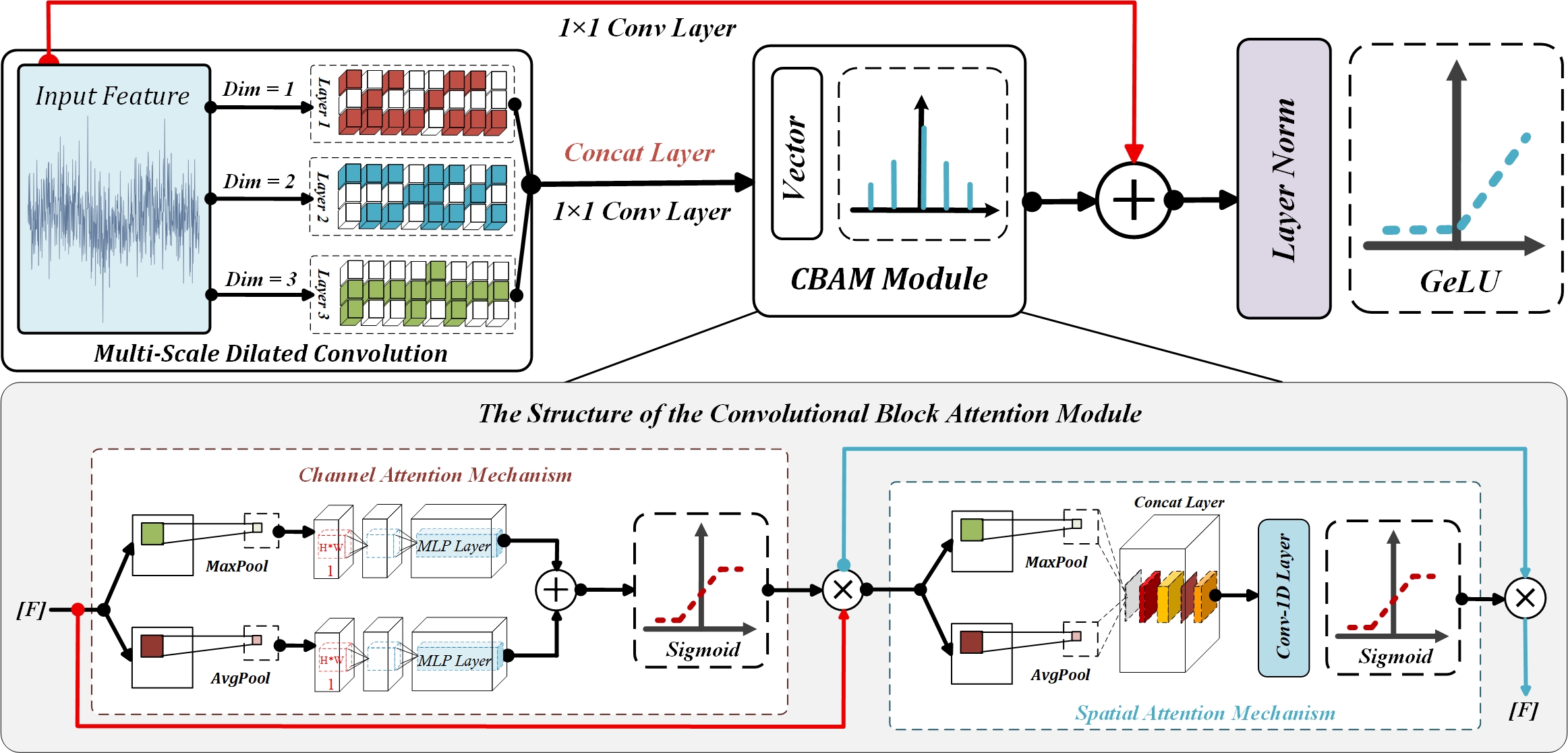}
	\caption{Architecture of the Proposed Convolutional Pre-Denoising Module with Multi-Scale Dilated Convolutions and CBAM-Based Attention Refinement.}
	\label{fig4}
\end{figure}

\vspace {-\baselineskip}
\subsection{Network architecture}  %
\textbf{Convolutional Pre-Denoising Module:} To improve the robustness of the proposed framework under practical LEO conditions, a Convolutional Pre-Denoising (CPD) module is deployed as the front end of the network, as illustrated in Fig. \ref{fig4}. Instead of directly feeding raw temporal features into the Transformer encoder, the CPD module first performs adaptive local feature refinement within the time domain.

Specifically, the module adopts a multi-scale parallel convolution architecture. To capture the temporal phase evolution under large Doppler shifts without introducing excessive computational overhead, standard kernel convolutions are replaced with $1 \times 3$ dilated convolutions using dilation rates of 1, 2, and 3. The three parallel branches therefore cover different temporal spans, allowing the module to model phase dependencies at local, intermediate, and extended ranges. The extracted feature maps are then concatenated and fused through a $1 \times 1$ convolution layer, which facilitates cross-channel information exchange and reduces the feature dimension.

To further enhance robustness against dynamic channel impairments, a Convolutional Block Attention Module (CBAM) is integrated following the feature fusion. Let $\mathbf{F}_{in} \in \mathbb{R}^{K \times N}$ denote the fused feature map input to the CBAM. By sequentially inferring attention weights along both the channel and temporal dimensions, the output of the CBAM, denoted as $\mathbf{F}_{cbam}$, is formulated as
\begin{equation}
	\label{CBAM}
	\mathbf{F}_{cbam} = \mathbf{M}_s(\mathbf{M}_c(\mathbf{F}_{in}) \otimes \mathbf{F}_{in}) \otimes (\mathbf{M}_c(\mathbf{F}_{in}) \otimes \mathbf{F}_{in})
\end{equation}

where $\mathbf{M}_c(\cdot)$ and $\mathbf{M}_s(\cdot)$ denote the weight generation functions of the channel attention module and the temporal (spatial) attention module, respectively. The symbol $\otimes$ represents element-wise multiplication. This mechanism dynamically emphasizes features sensitive to the CFO (such as phase and correlation attributes) while effectively suppressing irrelevant or heavily corrupted signal segments.

To preserve intrinsic waveform characteristics and ensure stable gradient propagation during training, a residual connection is incorporated. A $1 \times 1$ projection is applied to the initial shortcut branch to yield the residual feature $\mathbf{F}_{res}$, ensuring dimensional matching. Finally, to align the feature distribution with the mathematical manifold expected by the downstream self-attention mechanism, the aggregated representation is processed through Layer Normalization (LayerNorm) and a Gaussian Error Linear Unit (GELU) activation. The final output of the CPD module, $\mathbf{H}_{CPD} \in \mathbb{R}^{K \times N}$, is rigorously expressed as
\begin{equation}
	\label{CPD-out}
	\mathbf{H}_{CPD} = \text{GELU}(\text{LayerNorm}(\mathbf{F}_{cbam} \oplus \mathbf{F}_{res}))
\end{equation}

where $\oplus$ denotes element-wise addition. The GELU activation function introduces smooth non-linearity and is defined as $\text{GELU}(x) = x \Phi(x)$, where $\Phi(x)$ denotes the cumulative distribution function of the standard normal distribution. The refined latent sequence $\mathbf{H}_{CPD}$ is then directly fed into the subsequent Transformer encoder.

\begin{figure}[!t] 
	\centering
	\includegraphics[width=3.3in]{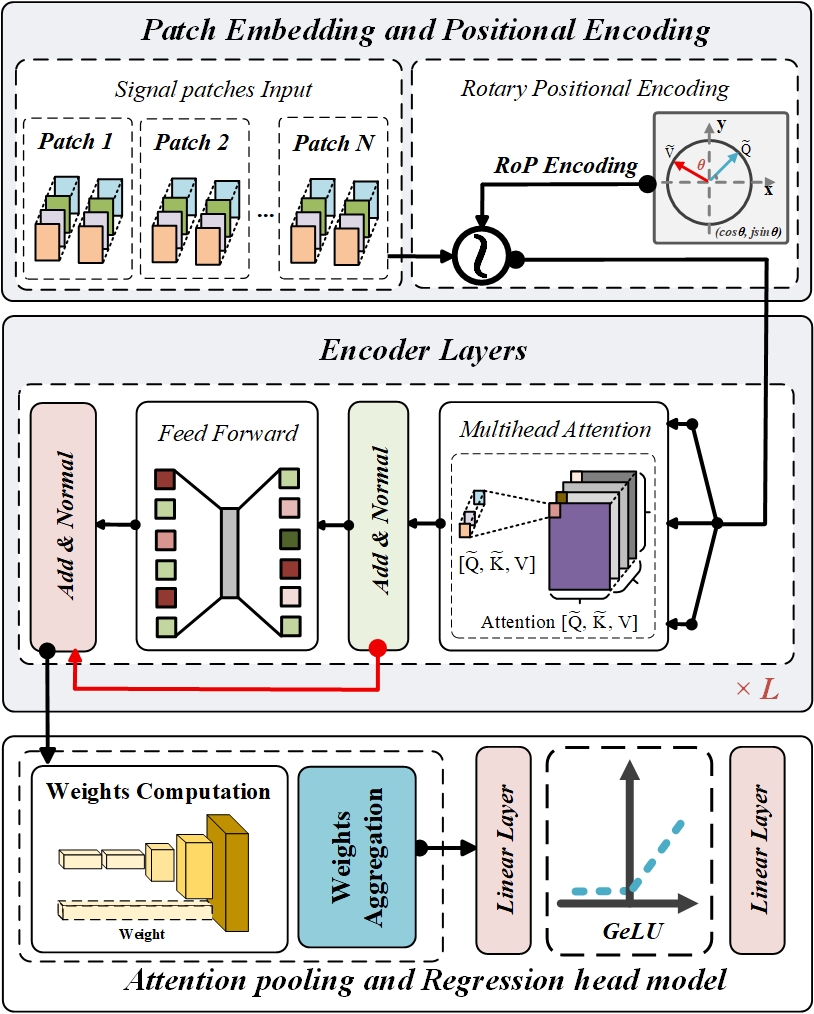}
	\caption{The architecture of the proposed Transformer encoder equipped with Rotary Positional Encoding and Attention Pooling.}
	\label{fig5}
\end{figure}

\textbf{Transformer Encoder Block:} After the CPD module refines the local temporal features, the processed latent sequence is forwarded to the Transformer-based modeling stage to explicitly capture long-range phase dependencies.

\subsubsection{Linear Embedding and Transformer Layers}
To transform the denoised features into a latent space suitable for the attention mechanism, a linear projection layer maps the sequence into an embedding dimension $d$:
\begin{equation}
	\mathbf{H}^{(0)} = \mathbf{H}_{CPD}^T \mathbf{W}_{e} + \mathbf{b}_{e}
\end{equation}

where $\mathbf{H}^{(0)} \in \mathbb{R}^{N \times d}$, $\mathbf{W}_{e} \in \mathbb{R}^{K \times d}$ is the projection matrix, and $\mathbf{b}_{e} \in \mathbb{R}^{d}$ is the bias. This embedded sequence is then processed by a stack of $L$ identical Transformer encoder layers.

\subsubsection{Rotary Positional Encoding (RoPE) and Attention Mechanism}
In the estimation of the CFO, the frequency offset physically manifests as a continuous phase rotation over time. Conventional absolute positional encodings, which rely on additive operations, struggle to inherently capture this multiplicative rotational geometry. Therefore, we explicitly employ the Rotary Positional Embedding (RoPE)\cite{RoPE} within the Multi-Head Self-Attention (MHSA) block.

Given the input representation $\mathbf{H}^{(l-1)}$ at the $l$-th layer, the mechanism first projects it into query ($\mathbf{Q}$), key ($\mathbf{K}$), and value ($\mathbf{V}$) matrices using learnable weights $\mathbf{W}_{Q}, \mathbf{W}_{K}, \mathbf{W}_{V} \in \mathbb{R}^{d \times d_h}$. For the $m$-th temporal position, RoPE strictly applies a block-diagonal rotation matrix $\mathbf{R}_{\Theta, m}$ to the corresponding query and key vectors $\mathbf{q}_m$ and $\mathbf{k}_m$
\begin{equation}
	\left\{
	\begin{aligned}
		\tilde{\mathbf{q}}_m &= \mathbf{q}_m \mathbf{R}_{\Theta,m} \\
		\tilde{\mathbf{k}}_m &= \mathbf{k}_m \mathbf{R}_{\Theta,m}
	\end{aligned}
	\right.
\end{equation}

where $\mathbf{R}_{\Theta, m}$ consists of $2 \times 2$ rotation matrices with rotation angles scaled by the position index $m$. By executing this geometric rotation in the latent space, the inner product of the query and key inherently encapsulates their relative temporal distance: $\tilde{\mathbf{q}}_m \tilde{\mathbf{k}}_n^T = \mathbf{q}_m \mathbf{R}_{\Theta, m-n} \mathbf{k}_n^T$. This mathematical property is structurally analogous to the progressive phase accumulation caused by the CFO, thereby explicitly equipping the self-attention mechanism with a strong inductive bias for tracking frequency offsets.

Subsequently, the rotated queries and keys are utilized to compute the attention weights, and the values are aggregated to form the output. For a single attention head, this operation is formulated as:
\begin{equation}
	\text{Attention}(\tilde{\mathbf{Q}}, \tilde{\mathbf{K}}, \mathbf{V}) = \text{softmax}\left(\frac{\tilde{\mathbf{Q}} \tilde{\mathbf{K}}^T}{\sqrt{d_h}}\right)\mathbf{V}
\end{equation}

where $d_h$ denotes the dimension of the head. The outputs from all parallel heads are concatenated and linearly projected to generate the refined sequence $\mathbf{H}_{attn}^{(l)}$. 

Following the attention sub-layer, a position-wise nonlinear transformation is applied independently at each temporal position:
\begin{equation}
	\mathbf{H}^{(l)} = \text{LayerNorm}\left(\mathbf{H}_{attn}^{(l)} + \text{Dropout}(\sigma(\mathbf{H}_{attn}^{(l)}\mathbf{W}_{1})\mathbf{W}_{2})\right)
\end{equation}

where $\mathbf{W}_{1}$ and $\mathbf{W}_{2}$ represent the learnable weight matrices of the linear transformations, and $\sigma(\cdot)$ denotes the GELU activation function. This iterative refinement process is repeated across all $L$ stacked encoder layers, progressively yielding a deep, globally contextualized representation of the input signal, which is subsequently forwarded to the downstream regression module.

\subsubsection{Attention Pooling and Regression Head}
In practical LEO satellite communication systems, the received signal is affected by rapidly varying distortions caused by high orbital mobility, severe Doppler effects, and noise contamination. As a result, the reliability of temporal features may vary significantly across the received sequence. Conventional aggregation strategies such as Global Average Pooling assign identical importance to all time steps, and therefore may be sensitive to distorted observations, leading to suboptimal global representations.

To address this issue, we introduce an Attention Pooling mechanism to dynamically evaluate the reliability of each temporal feature. Given the final contextualized sequence $\mathbf{H}^{(L)}$, a learnable linear transformation followed by a softmax function is utilized to compute the attention weight vector $\mathbf{a} \in \mathbb{R}^{N}$:
\begin{equation}
	\mathbf{a} = \text{softmax}(\mathbf{H}^{(L)} \mathbf{W}_{attn})
\end{equation}

where $\mathbf{W}_{attn} \in \mathbb{R}^{d \times 1}$ maps the high-dimensional features into scalar weights. The aggregated, noise-resistant representation $\mathbf{h}_{pool} \in \mathbb{R}^{d}$ is then obtained by computing the weighted sum across the temporal dimension:
\begin{equation}
	\mathbf{h}_{pool} = \sum_{n=1}^{N} a_n \mathbf{h}_n^{(L)}
\end{equation}

where $a_n$ and $\mathbf{h}_n^{(L)}$ denote the $n$-th element of $\mathbf{a}$ and the $n$-th row vector of $\mathbf{H}^{(L)}$, respectively. This soft-selection mechanism explicitly enables the network to emphasize reliable, high-SNR phase features while ignoring severely distorted segments.

Finally, $\mathbf{h}_{pool}$ is fed into a multi-layer fully connected network to predict the estimated normalized CFO $\hat{\varepsilon}$:
\begin{equation}
	\hat{\varepsilon} = \mathbf{W}_{FC2} \sigma \left( \mathbf{W}_{FC1} \mathbf{h}_{pool} + \mathbf{b}_{FC1} \right) + b_{FC2}
\end{equation}

where $\mathbf{W}_{FC1} \in \mathbb{R}^{d_{hidden} \times d}$ and $\mathbf{W}_{FC2} \in \mathbb{R}^{1 \times d_{hidden}}$ denote the weight matrices, and $\mathbf{b}_{FC1}$ and $b_{FC2}$ are the corresponding biases. By embedding the structural prior of temporal phase evolution directly into the attention mechanism, the proposed architecture effectively captures the underlying CFO dynamics, thereby ensuring high-precision frequency offset estimation even under severe impairments.

\vspace {-\baselineskip}
\subsection{Loss Function}
In deep learning-based CFO estimation, the MSE is commonly adopted as the default loss function for regression tasks. However, in LEO satellite communication scenarios, the received signal may undergo strong time variation and severe Doppler-induced distortion, which can lead to samples with relatively large estimation errors during training. Because the MSE penalizes errors quadratically, it is highly sensitive to such deviations, which may adversely affect training stability.

To mitigate this issue and improve the robustness of the proposed network, we adopt the Huber loss function for model training. The Huber loss integrates the benefits of both the mean absolute error (MAE) and the MSE. When the estimation error is large, it behaves as an L1 loss, which bounds the gradient magnitude and reduces the sensitivity to outliers. When the error is small, it smoothly transitions to an L2 loss, enabling fine-grained convergence near the optimum.

Given a mini-batch of $B$ training samples, let $\varepsilon_i$ denote the ground-truth normalized CFO and $\hat{\varepsilon}_i$ denote the estimated CFO generated by the network for the $i$-th sample. The Huber loss is formulated as:
\begin{equation}
	\mathcal{L}_{Huber} = \frac{1}{B} \sum_{i=1}^{B} l_{i}(\varepsilon_i, \hat{\varepsilon}_i)
\end{equation}

where the instance loss $l_i$ is defined by a threshold parameter $\delta$ as:
\begin{equation}
	l_{i} = 
	\begin{cases} 
		\frac{1}{2}(\varepsilon_i - \hat{\varepsilon}_i)^2, & \text{if } |\varepsilon_i - \hat{\varepsilon}_i| \le \delta \\
		\delta |\varepsilon_i - \hat{\varepsilon}_i| - \frac{1}{2}\delta^2, & \text{otherwise} 
	\end{cases}
\end{equation}

The hyperparameter $\delta$ controls the transition point between the L1 and L2 penalty regions. During network training, the proposed Convolution-Augmented Transformer is optimized using the Huber loss, which improves robustness to large estimation deviations and stabilizes convergence.

\vspace {-\baselineskip}
\subsection{Computation Complexity}
To further elucidate the properties of the proposed model, we analyze the computational complexity based on the underlying tensor operations of each module. The network fundamentally comprises three primary stages: the Convolutional Pre-Denoising (CPD) module, the Transformer encoder block, and the final output head, which encompasses Attention Pooling and a Regression Head. Let $N$ denote the sequence length, $C_{in}$ represent the dimension of the input channel, and $d$ indicate the dimension of the hidden embedding. The following analysis first considers the computational complexity for a single input sample, and the batch complexity can be obtained by further multiplying by the batch size $B$. Accordingly, we conduct a separate analysis for each component. 

The CPD module relies on multi-scale dilated 1D convolutions, intermediate feature fusion, and a Convolutional Block Attention Module (CBAM). The computational complexity of the parallel multi-scale convolution branches and the subsequent $1 \times 1$ fusion layer is formulated as:
\begin{equation}
	\mathcal{O}(N d (C_{in} K + d))
\end{equation}

where $K$ denotes the kernel size of the dilated convolutions. The channel and spatial attention mechanisms within the CBAM introduce a minor overhead of $\mathcal{O}(d^2 / r + N)$, where $r$ denotes the reduction ratio of the multi-layer perceptron.

The primary computational bottleneck of the proposed architecture resides within the Transformer encoder. For $L_e$ stacked Transformer layers, the time complexity encompasses the Multi-Head Self-Attention (MHSA) mechanism and the position-wise Feed-Forward Network (FFN). The MHSA requires $\mathcal{O}(N^2 d)$ operations for the sequence-wise dot products and $\mathcal{O}(N d^2)$ operations for the linear projections. The FFN, which expands and compresses the hidden dimension, requires $\mathcal{O}(N d^2)$ operations. Thus, the computational complexity of the Transformer encoder block is expressed as:
\begin{equation}
	\mathcal{O}(L_e N d (N + d))
\end{equation}

Finally, the output stage incorporates Attention Pooling to aggregate the global temporal features, alongside a multi-layer Regression Head to predict the final frequency offset. These operations involve linear mappings over the temporal dimension and the hidden dimension, contributing an additional computational complexity of $\mathcal{O}(N d + d^2)$.

To provide a comprehensive overview, Table \ref{tab:complexity} summarizes the theoretical time complexity of the proposed architecture. Components with minor computational contributions, such as layer normalization, activation functions, and RoPE operations, are collectively denoted by $D_T$.

\begin{table}[!t]
	\centering
	\caption{Computational Complexities of the Proposed Network}
	\label{tab:complexity}
	\renewcommand{\arraystretch}{1.5}
	\begin{tabular}{cc}
		\toprule
		\textbf{Module Name} & \textbf{Computational Complexity} \\
		\midrule
		CPD Module & $\mathcal{O} \left( B N d (C_{in} K + d) \right)$ \\
		Transformer Encoder & $\mathcal{O} \left( B L_e N d (N + d) \right)$ \\
		Output Head & $\mathcal{O} \left( B (N d + d^2) \right)$ \\
		\midrule
		\textbf{Overall} & $\mathcal{O} \left( B N d (C_{in} K + d + L_e(N + d)) + B D_T \right)$ \\
		\bottomrule
	\end{tabular}
\end{table}
\begin{table}[htbp]
	\centering
	\caption{Simulation Parameters}
	\label{tab:simulation_parameters}
	\begin{tabular}{lc}
		\toprule
		\textbf{Parameter} & \textbf{Value} \\
		\midrule
		Carrier frequency & $2.6\,\mathrm{GHz}$ \\
		Speed of light &  $3 \times 10^8\,\mathrm{m/s}$\\
		Relative velocity & $6000$--$8000\,\mathrm{m/s}$ \\
		Signal length & $64 / 128 / 256$ \\
		SNR range & $0$--$15\,\mathrm{dB}$  \\
		Modulation types & QPSK 	\\
		Number of samples & $100000$ \\
		Training:Validating:Testing & $6:2:2$ \\
		\bottomrule
	\end{tabular}
\end{table}

\section{SIMULATION RESULTS AND ANALYSIS}
This study constructs a comprehensive dataset to evaluate the proposed Convolution-Augmented Transformer architecture for CFO estimation within OFDM communication systems operating in LEO satellite networks. The dataset comprises 100,000 QPSK-modulated signals with varying symbol lengths of 64, 128, and 256, transmitted over an AWGN channel. These signals are generated under various SNR conditions ranging from 0 dB to 15 dB with a step size of 3 dB. The carrier frequency is set to 2.6 GHz, which is representative of typical satellite communication bands, whereas the relative velocities between the satellite and the ground terminal range from 6000 m/s to 8000 m/s, corresponding to realistic orbital characteristics of LEO satellites. These velocity ranges induce dynamic Doppler shifts ranging from approximately 50 kHz to 70 kHz. To facilitate the training of the neural network and ensure stable gradient propagation, the Doppler shift $f_d$ is explicitly normalized and mapped to a scaled label $\bar{\varepsilon} \in [-0.5, 0.5]$ via normalization. Specifically, the mapping is formulated as:
\begin{equation}
	\label{eq:norm_cfo}
	\bar{\varepsilon} = \frac{f_d - f_{\min}}{f_{\max} - f_{\min}} - 0.5
\end{equation}

where $f_{\min}$ and $f_{\max}$ denote the minimum and maximum Doppler frequency offsets considered within the generated dataset, respectively. The detailed system parameters and specifications of the dataset are summarized in Table \ref{tab:simulation_parameters}.

\begin{table}[htbp]
	\centering
	\caption{Training Parameters Setting}
	\label{tab:training_params}
	\begin{tabular}{cc}
		\toprule
		\textbf{Parameters} & \textbf{Values} \\
		\midrule
		Batch size & 256 \\
		Numbers of epoch & 200 \\
		Learning rate & 0.001 \\
		Optimizer & Adam \\
		\bottomrule
	\end{tabular}
\end{table}
Table \ref{tab:training_params} presents the detailed training parameters of the network. The dataset was split into training, validation, and test sets in a 6:2:2 ratio, corresponding to $60\%$, $20\%$, and $20\%$ of the samples, respectively. This split supports model tuning while preserving an unbiased evaluation on held-out data. All simulations and network training were carried out on an NVIDIA GeForce RTX 3090 GPU for efficient computation. Although the model was optimized with the Huber loss, mean squared error (MSE) was used as the main evaluation metric to measure the deviation between the estimated and true normalized CFO values.

\subsection{Ablation Study on Input Representations}

\begin{figure}[htbp]
	\centering
	\includegraphics[width=0.5\textwidth]{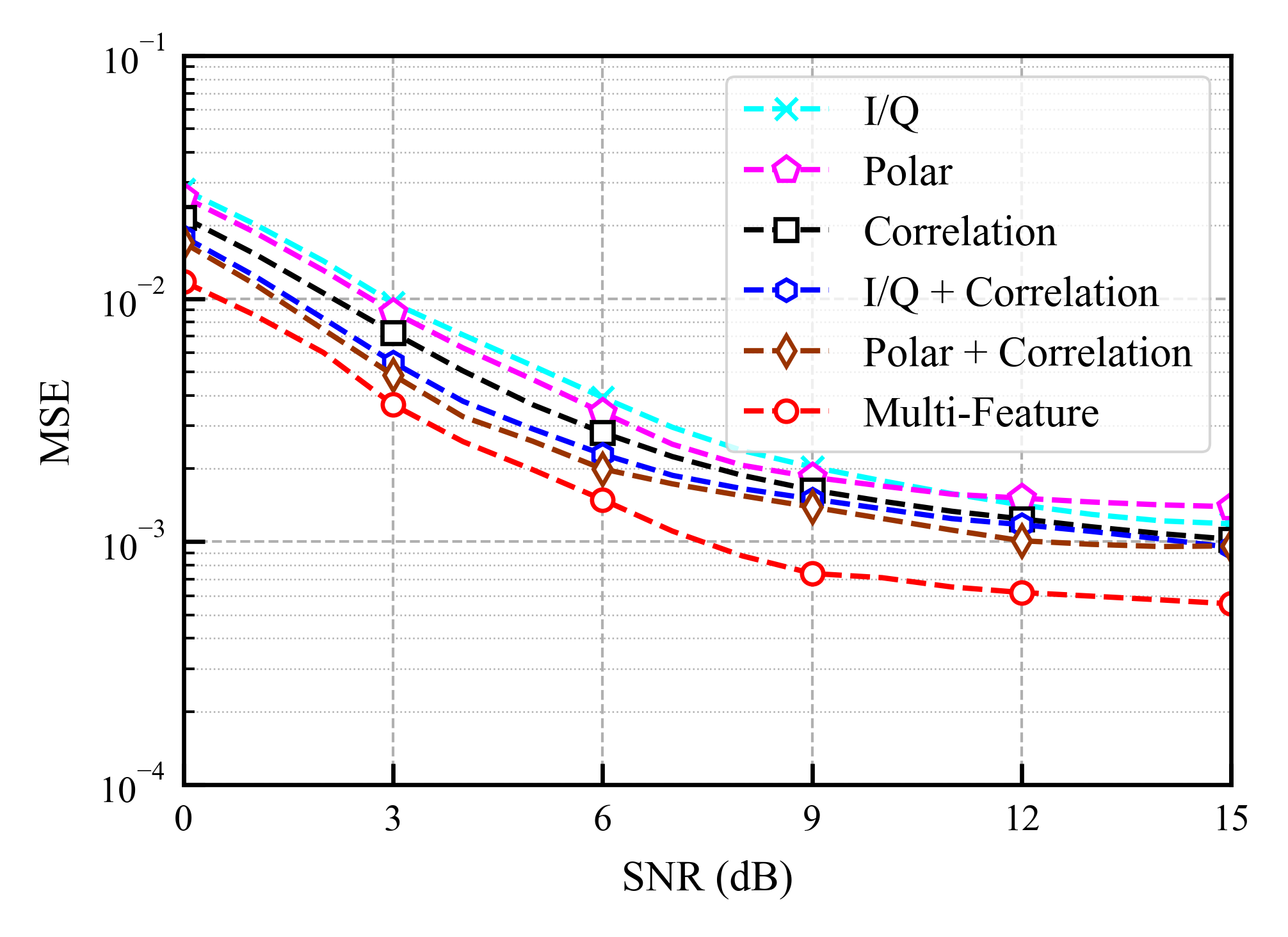}
	\caption{Performance Comparison of CFO Estimation Using Different Input Feature Combinations Across Varying SNRs.}
	\label{fig:res_input}
\end{figure}

To validate the effectiveness of the proposed multi feature construction, Fig.~\ref{fig:res_input} illustrates the CFO estimation performance under various input feature combinations. Across the entire evaluated SNR range from 0 dB to 15 dB, the proposed multi feature approach consistently yields the lowest estimation error. Specifically, under severe noise conditions at 0 dB, the proposed configuration achieves a substantial performance gain, reducing the MSE by approximately 58.1\% relative to the raw I/Q input. This performance advantage persists even at a high SNR of 15 dB, maintaining an error reduction of over 53.1\%. Fundamentally, this improvement can be mainly attributed to the strong information complementarity introduced by the explicit fusion of short-term instantaneous physical attributes and long-term statistical properties. By integrating these heterogeneous yet complementary features, the proposed architecture effectively alleviates the learning burden of the self-attention mechanism in capturing dynamic Doppler shifts.

The ablation results further reveal that the incorporation of statistical correlation is the primary driver of this performance enhancement. The correlation feature alone demonstrates robust noise resistance, achieving an error reduction of approximately 23.0\% at 0 dB, noticeably outperforming the raw I/Q baseline. Moreover, combining correlation data with physical inputs systematically improves estimation fidelity. For instance, fusing correlation with I/Q data reduces the error by nearly 36.0\% at 0 dB, while the polar-correlation fusion further decreases it by 39.5\%. This improvement arises because the normalized correlation function deemphasizes uncorrelated noise while highlighting the periodic phase variations induced by the Doppler effect. Consequently, it provides highly reliable statistical priors, enabling the attention mechanism to accurately identify periodic deviations even within heavily corrupted signals.

\subsection{CFO Estimation performance}

\begin{figure}[htbp]
	\centering
	\includegraphics[width=0.4\textwidth]{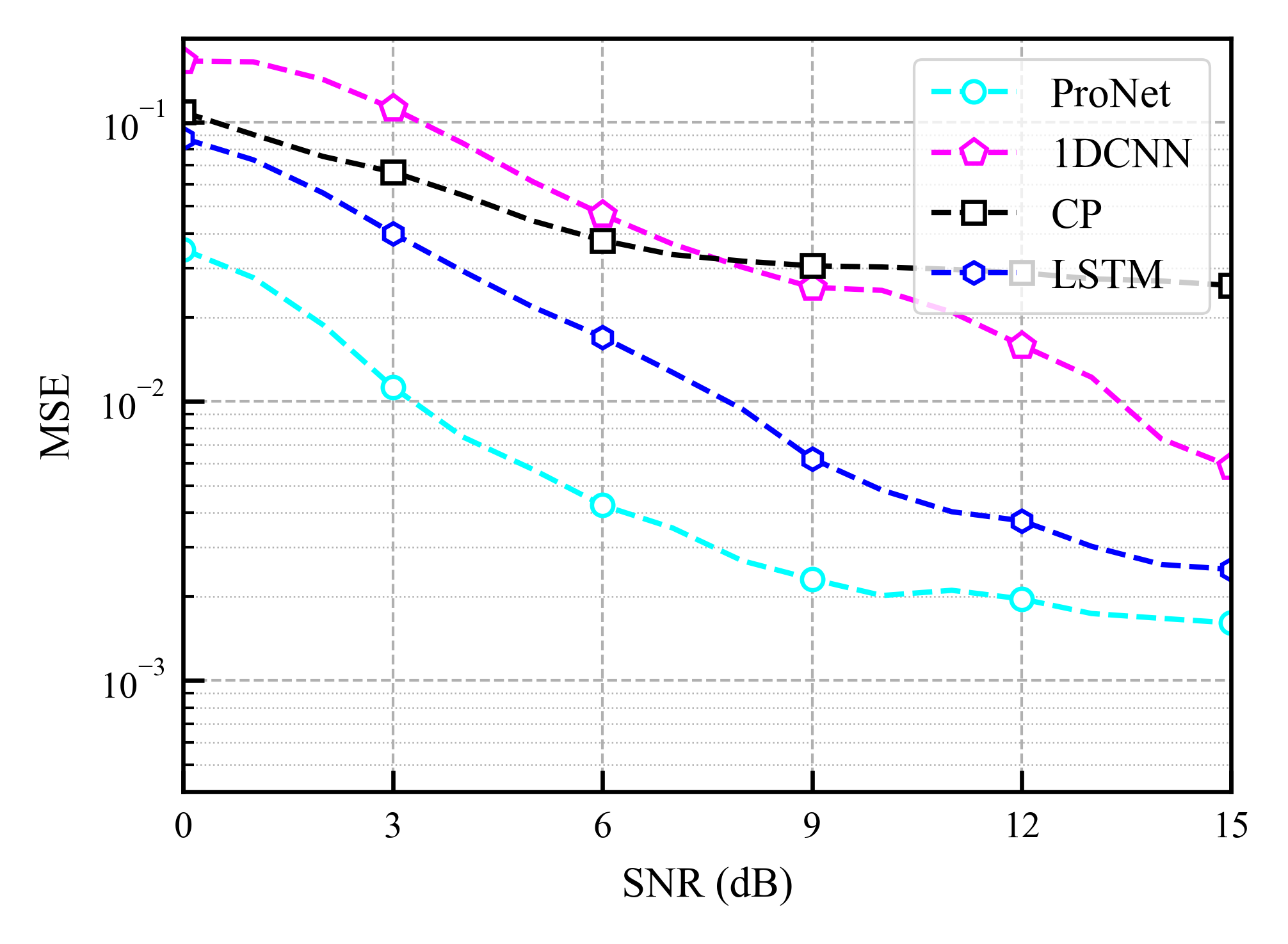}
	\caption{
		MSE Performance Comparison Across Different Simulation Methods with an Input Signal Length of 64}
	\label{fig:dif_method_64}
\end{figure}

\begin{figure}[htbp]
	\centering
	\includegraphics[width=0.4\textwidth]{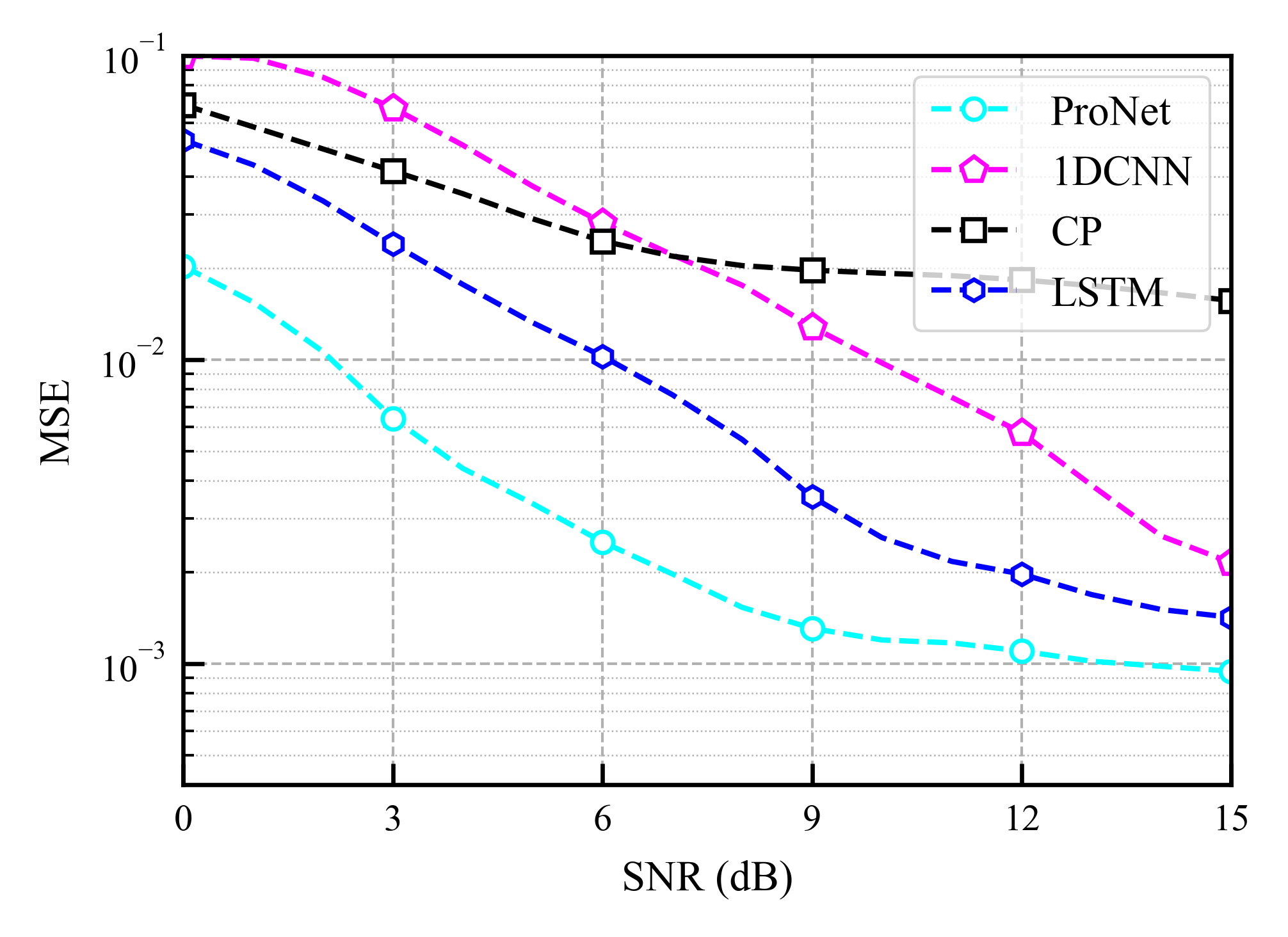}
	\caption{
		MSE Performance Comparison Across Different Simulation Methods with an Input Signal Length of 128}
	\label{fig:dif_method_128}
\end{figure}

\begin{figure}[htbp]
	\centering
	\includegraphics[width=0.4\textwidth]{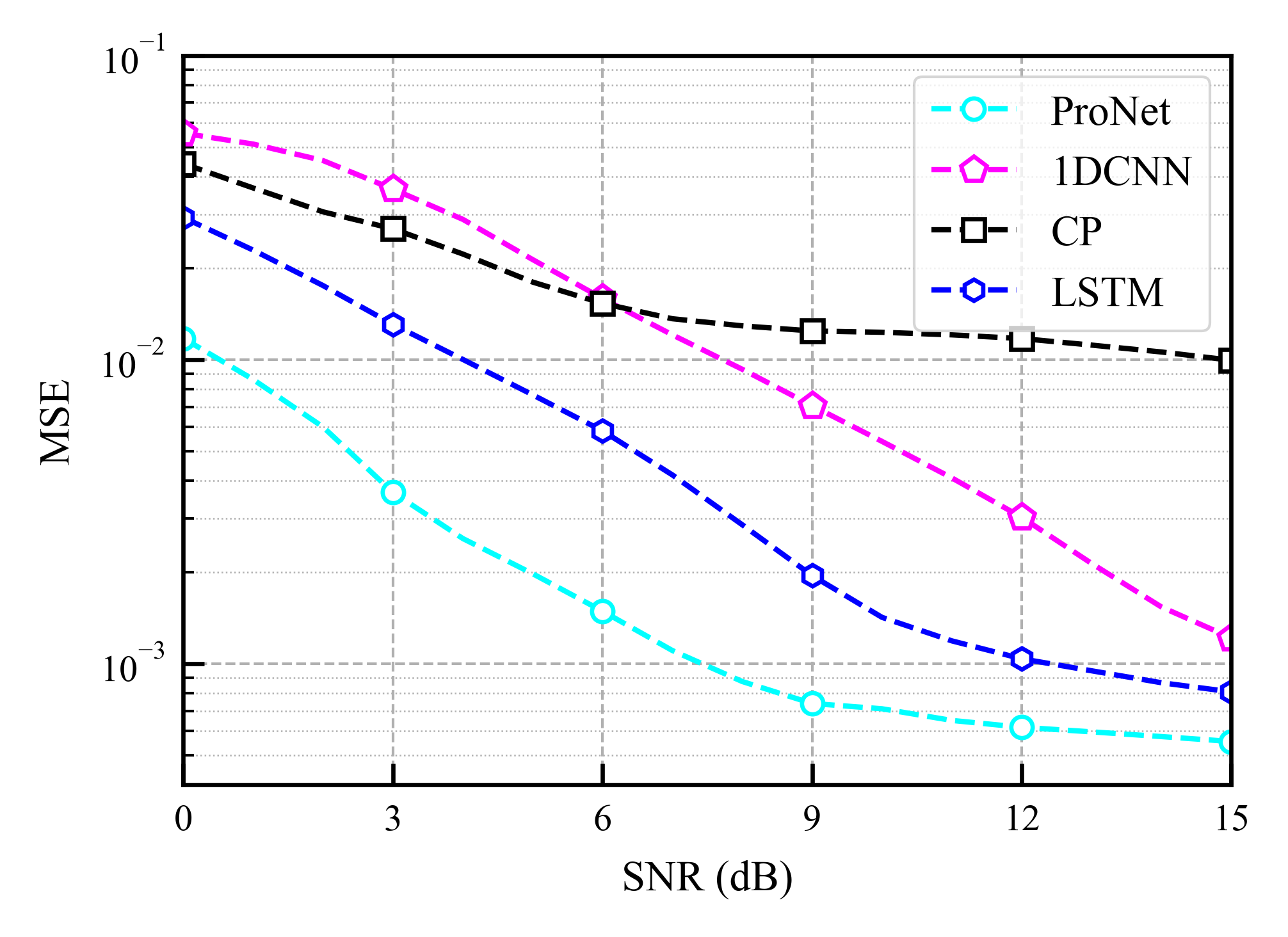}
	\caption{
		MSE Performance Comparison Across Different Simulation Methods with an Input Signal Length of 256}
	\label{fig:dif_method_256}
\end{figure}

To verify the performance of the proposed CFO estimation method, we compare the approach with existing methods, including the CP correlation method \cite{cp-2}, CFO estimation based on the 1D-CNN network \cite{cfoeg1}, and CFO estimation based on the LSTM network \cite{rnn-2}. For a fair comparison, all methods are evaluated on the dataset generated according to the parameters detailed in Table \ref{tab:simulation_parameters}, and the hyperparameters are set to the configurations reported in the respective literature. Fig.~\ref{fig:dif_method_64}, Fig.~\ref{fig:dif_method_128}, and Fig.~\ref{fig:dif_method_256} report the CFO estimation results for input signal lengths of 64, 128, and 256, respectively. Across all evaluated methods, expanding the signal length consistently reduces the mean squared error over the entire SNR range, confirming that a broader observation window provides more reliable phase-evolution cues for frequency offset estimation. This trend is particularly pronounced for the proposed network. With each doubling of the input sequence length, the estimation error of the proposed method is approximately halved under both low- and high-SNR conditions. This demonstrates a highly stable scaling property, ensuring that the proposed architecture remains the best-performing approach across all evaluated temporal scales.

As illustrated in Fig.~\ref{fig:dif_method_64}, which corresponds to the shortest input length, the proposed method achieves the lowest estimation error across the entire SNR range, demonstrating strong robustness even when the available temporal evidence is limited. At 0~dB, the proposed architecture reduces the estimation error by more than sixty percent compared with the LSTM network and achieves an even larger improvement over the conventional CP and convolutional methods. As the SNR increases to 15~dB, the accuracy difference between the proposed method and the other deep-learning baselines becomes smaller; nevertheless, the proposed network still maintains a clear advantage. Notably, the conventional CP method exhibits a pronounced error floor in the high-SNR regime, whereas the 1D-CNN model performs the worst in the low-SNR regime. This result suggests that purely local convolutions are insufficient for inferring complex frequency-offset dynamics from short and noisy observations.

When the observation length is extended to 128, as shown in Fig.~\ref{fig:dif_method_128}, all reference methods demonstrate improved accuracy with the increased sequence length, yet their relative performance ordering remains unchanged. The proposed method consistently sustains its superiority, outperforming the LSTM, 1D-CNN, and CP approaches across medium-to-high SNR levels. Although the absolute performance margins among the deep learning models narrow compared to the shortest sequence configuration, the proposed network continues to yield the lowest estimation error under all evaluated SNR conditions. This persistent advantage indicates that the integration of the convolutional pre-denoising module and the self-attention mechanism not only improves robustness under low-SNR conditions, but also substantially enhances the ability to capture long range phase dependencies as more temporal samples become available.

In the longest input configuration presented in Fig.~\ref{fig:dif_method_256}, all evaluated schemes achieve lower estimation error as the observation length is further extended. As the reference methods are also able to exploit additional phase information from the longer sequence, the proposed architecture still maintains a consistent performance advantage across both low- and high-SNR regions. One noticeable observation in this setting is that the CP correlation method benefits less from the extended observation length. This suggests that simply enlarging the observation window may not be sufficient to fully overcome the inherent limitations of short-range cyclic-prefix correlation in dynamic LEO scenarios. By comparison, the proposed data-driven model is able to make better use of the additional time-domain samples and achieve further error reduction. These results suggest that the explicit fusion of multi-view features, together with the global receptive field provided by the Transformer encoder, is beneficial for capturing fine-grained periodic phase variations from longer signal sequences.

\subsection{Impact of Loss Function on Training Robustness}
\begin{figure}[htbp]
	\centering
	\includegraphics[width=0.4\textwidth]{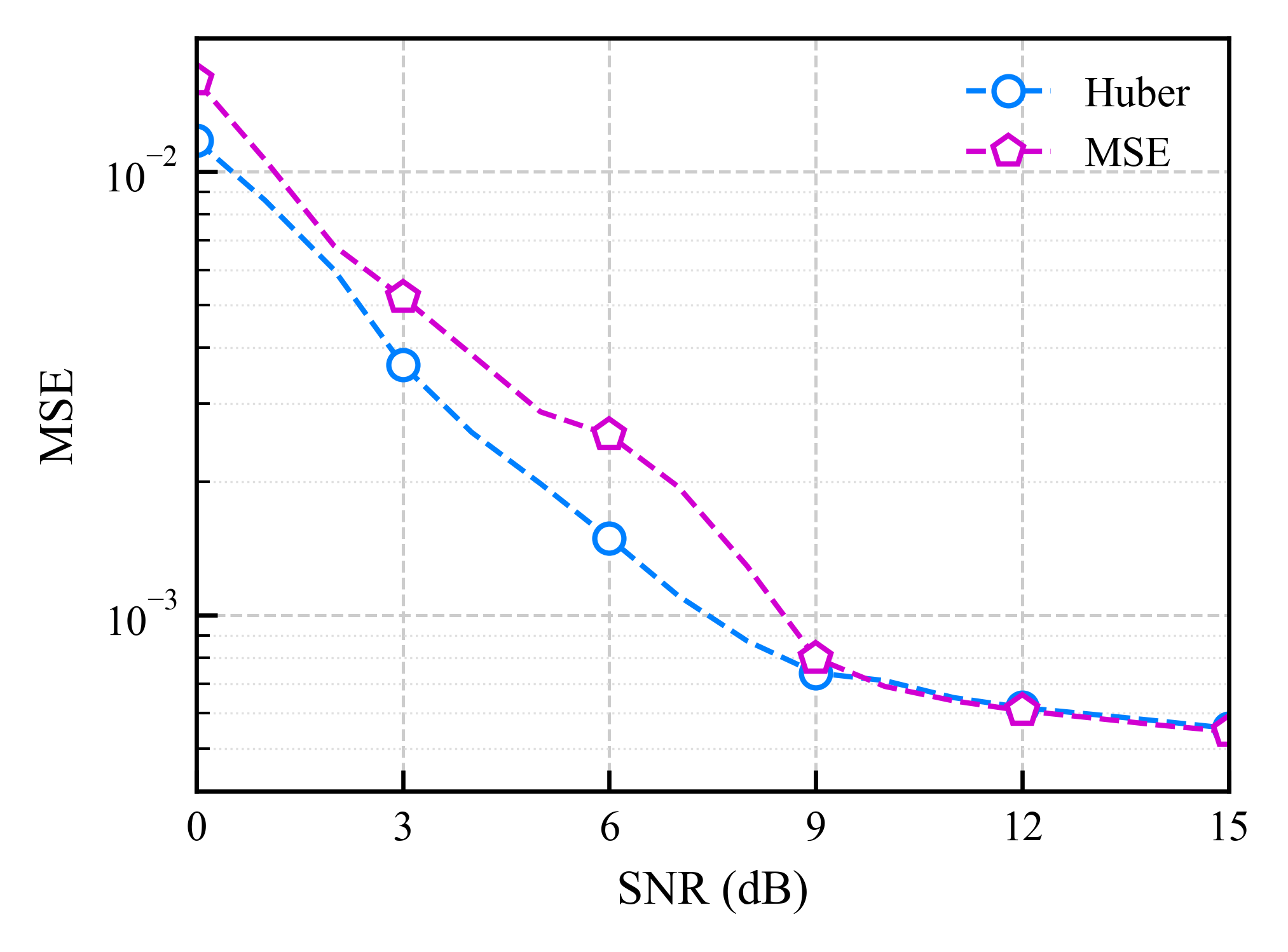}
	\caption{Performance Comparison of Huber Loss and MSE Loss under Different SNR Values.}
	\label{fig:HuberLoss}
\end{figure}

To assess the effectiveness of the robust training strategy, we further examine the estimation behavior of the proposed network under the Huber loss and the standard MSE loss. Fig.~\ref{fig:HuberLoss} illustrates the corresponding performance curves over the evaluated SNR range. It can be observed that, under low SNR conditions, the model trained with the Huber loss achieves lower estimation error. For instance, at 0~dB, the estimation error is reduced by approximately 26.8\% compared with that obtained using the MSE loss. This result suggests that the Huber loss is better suited to training under low SNR and highly dynamic conditions, where the received signal is more likely to be affected by strong disturbances and the resulting errors may exhibit more pronounced outlier characteristics. Because the MSE loss imposes a quadratic penalty on large errors, the optimization process can become more sensitive to these abnormal samples. By comparison, the Huber loss suppresses the influence of large deviations through its piecewise formulation, thereby improving training stability and robustness. When the SNR exceeds 9~dB, the performance gap between the two loss functions gradually narrows. This tendency is consistent with the mathematical property of the Huber loss. When the prediction error is small, the Huber loss behaves increasingly similarly to a quadratic penalty, and its optimization characteristic gradually approaches that of the MSE loss. Therefore, while the Huber loss provides better robustness in low SNR conditions, it also maintains competitive estimation performance in high-SNR regimes, where its behavior becomes close to that of the MSE loss.

\subsection{Impact of Positional Encoding Strategies}
\begin{figure}[htbp]
	\centering
	\includegraphics[width=0.4\textwidth]{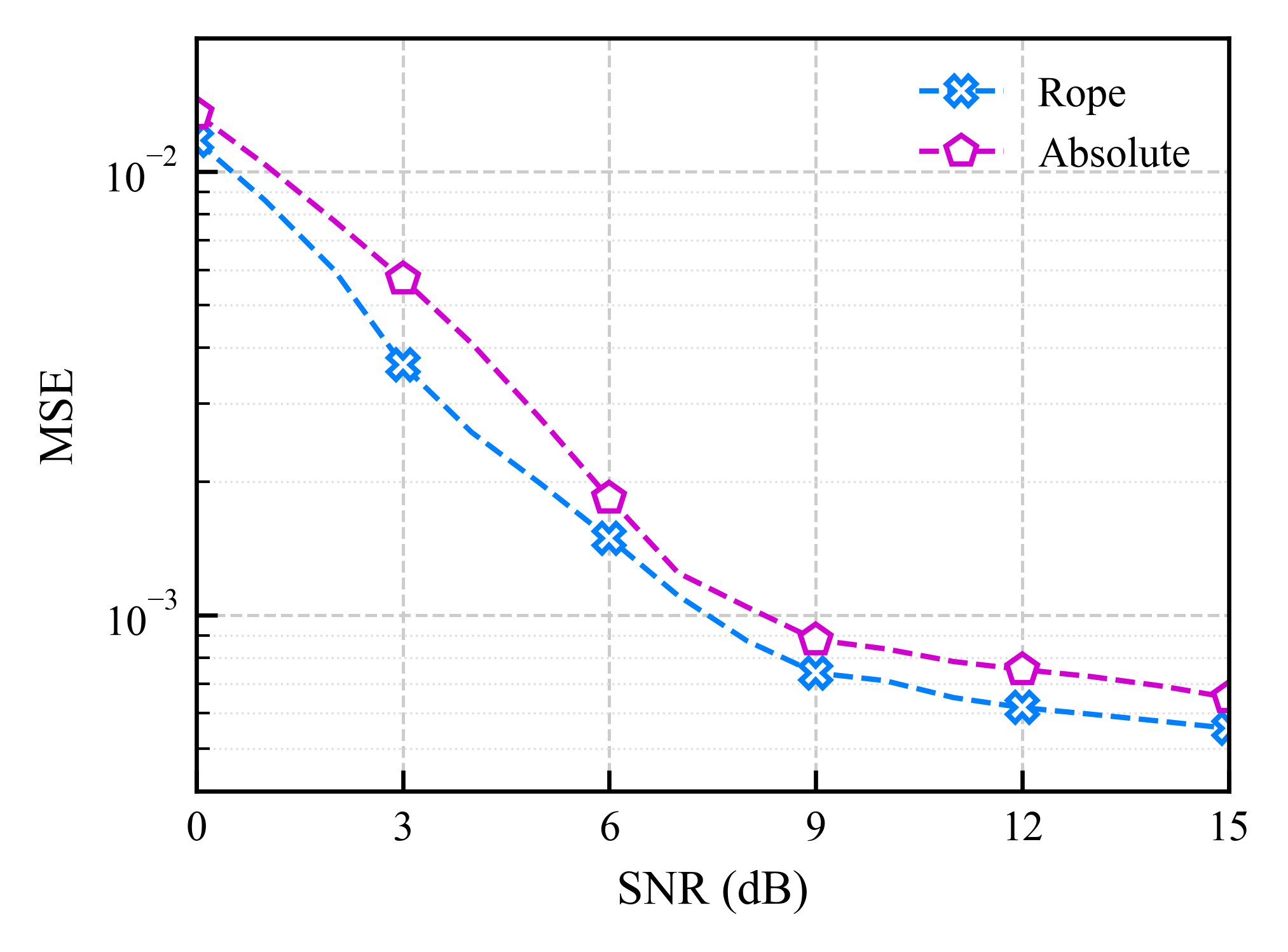}
	\caption{Comparison of MSE Performance Using Different Positional Encodings in the Proposed Transformer Network.}
	\label{fig:Rope}
\end{figure}

To examine the effect of positional encoding schemes on estimation accuracy, we compare the proposed architecture equipped with Rotary Positional Encoding (RoPE) with a baseline using standard absolute positional encoding. Fig.~\ref{fig:Rope} shows the MSE curves of the two encoding strategies across the evaluated SNR range. As shown in the figure, the RoPE-based network consistently achieves lower estimation error than the counterpart with absolute positional encoding under all tested conditions. At 0~dB, the use of RoPE reduces the estimation error by approximately 12.8\% relative to the baseline. This advantage becomes more evident as the signal quality improves, and it remains stable throughout the high SNR regime, with an error reduction of approximately 14.7\% still maintained at 15~dB. These results suggest that incorporating relative positional information through RoPE helps the self-attention mechanism better capture the underlying dependencies relevant to CFO estimation.

RoPE is well suited to this task because frequency offset in communication signals is reflected in the evolution of signal phase. In Doppler-affected OFDM systems, the carrier frequency offset manifests as a gradual phase rotation over time. Standard absolute positional encoding introduces positional information through additive embeddings, which are not well matched to this rotational pattern. By contrast, RoPE encodes positional information through rotations in the latent space, allowing the inner product of the query and key vectors to reflect relative phase variation over time more naturally. This property is more consistent with the temporal phase accumulation induced by dynamic frequency offset and therefore provides a more appropriate inductive bias for phase evolution modeling and Doppler shift tracking. As a result, the network can capture long-range temporal dependencies more effectively and achieve more accurate CFO estimation under both low- and high SNR conditions.

\section{CONCLUSIONS AND FUTURE WORK}

This paper proposes a Convolution-Augmented Transformer architecture to address the severe performance degradation of OFDM systems caused by massive Doppler shifts and fast time-varying fading in LEO satellite communications. To improve robustness in highly dynamic satellite channels, the proposed method constructs a multi-feature representation that jointly exploits I/Q components, polar-domain representations, and statistical correlation information, thereby providing complementary views of the received signal. On this basis, a Convolutional Pre-Denoising (CPD) module is introduced to suppress localized noise contamination and extract stable spatiotemporal patterns before global sequence modeling. The resulting features are then fed into a Transformer encoder equipped with RoPE and an attention pooling mechanism, enabling the network to model long-range phase dependencies and track dynamic frequency offsets more effectively. In addition, the Huber loss is adopted during training to reduce sensitivity to extreme channel outliers and improve optimization robustness. Extensive simulation results demonstrate that the proposed method consistently outperforms conventional model-based methods and representative deep learning approaches in terms of estimation accuracy and robustness across a wide range of challenging LEO channel conditions.

Despite these advantages, the proposed framework incurs a relatively high computational overhead due to the self-attention mechanism within the Transformer encoder, potentially limiting its practical deployment on resource-constrained platforms. Therefore, future work will focus on reducing model complexity via lightweight architectural designs—such as sparse attention, efficient token compression, and knowledge distillation—while preserving estimation accuracy. Another promising direction is to tightly integrate the CFO estimation module with downstream signal processing tasks, particularly channel decoding, to develop an end-to-end joint synchronization and decoding framework for highly dynamic satellite communication networks.

\bibliographystyle{IEEEtran}
\bibliography{ref}

\end{document}